\documentclass[conference]{IEEEtran}
\IEEEoverridecommandlockouts

\newcommand{\controllermodel}{FIXME: controller model, vendor and version}
\newcommand{\reportmodel}{FIXME: report model, vendor and version}
\usepackage{cite}
\usepackage{amsmath,amssymb,amsfonts}
\usepackage{algorithm}
\usepackage{algpseudocode}
\usepackage{graphicx}
\usepackage{textcomp}
\usepackage{booktabs}
\usepackage{url}
\usepackage[hidelinks]{hyperref}
\usepackage{balance}

\graphicspath{{figures/}}

\def\BibTeX{{\rm B\kern-.05em{\sc i\kern-.025em b}\kern-.08em
    T\kern-.1667em\kern-.125emX}}

\begin{document}

\title{Agentic Cloud Decoys: \\ A Deception-Driven Framework for\\ Autonomous Intrusion Investigation}

\author{\IEEEauthorblockN{Mohan Kumar Manivannan}
\IEEEauthorblockA{\textit{College of Information Science} \\
\textit{University of Arizona}\\
Tucson, AZ, USA \\
mohanmanivannan@arizona.edu}
\and
\IEEEauthorblockN{Dalal N. Alharthi}
\IEEEauthorblockA{\textit{College of Information Science} \\
\textit{University of Arizona}\\
Tucson, AZ, USA \\
dalharthi@arizona.edu}
}

\maketitle

\begin{abstract}
Cloud environments produce control plane and data plane telemetry at a scale that, paradoxically, makes intrusion understanding harder rather than easier. Attackers operate through legitimate identity, federated session tokens, and cloud native APIs whose surface signature is indistinguishable from routine administrative behavior, and analysts spend most of an incident reconstructing context that the logs technically already contain. This paper presents Cloud Decoy AI Agent, a deception driven investigation framework that pairs a high fidelity cloud decoy with an autonomous large language model agent in order to compress the path from suspicious activity to analyst ready incident report. We argue that connecting a decoy to an agent is not a wiring exercise, and we identify three properties of cloud telemetry that make it a design problem. The unit of investigation is the session rather than the event, and the correct session key is obscured by the identity layering that federated credentials introduce. The agent's evidence horizon has to be bounded, because an agent permitted to query the full control plane history inherits both the cost profile and the false positive profile that deception was introduced to remove. And cloud telemetry is partly adversary authored, since object keys and user agent strings are attacker chosen values that providers record verbatim, which turns any log to prompt path into an indirect prompt injection channel that a decoy widens rather than narrows. We address the first two with a formal session aggregation operator over a four element pivot tuple drawn only from provider derived fields, and with dynamic prompt generation, a two stage prompt assembly strategy that enforces a stated grounding invariant by carrying only fields the agent actually observed. We identify the third as an unaddressed exposure in this class of system, specify the mitigation it requires, and are explicit that the prototype evaluated here does not implement it. Evaluation covers ten controlled scenarios on AWS S3 spanning enumeration, retrieval, multi object access, denied access, destructive deletion, delayed access, and script driven variants. Nine of ten scenarios were reconstructed completely, no report contained an assertion the authors could not trace to an observed artifact, and end to end latency was four to five minutes under a configuration whose flush window dominates that figure. We state plainly what this evaluation does not establish: it has no baseline, no ablation, no independent scoring, and one service on one provider, and we name the specific comparisons that would settle each question.
\end{abstract}


% =====================================================================
\section{Introduction}
% =====================================================================
Public cloud platforms now operate at a scale where the unit of telemetry is the API call rather than the packet, and where the boundary between legitimate and malicious activity is rarely visible at the layer of any single event. A practitioner inspecting CloudTrail or its equivalent on any major provider sees thousands of records per hour, almost all of them benign, almost all of them associated with valid identities. When an attacker reaches the control plane through credential compromise, federated session abuse, or misconfigured trust, the actions they perform, such as \texttt{ListObjects}, \texttt{GetObject}, and \texttt{GetCallerIdentity}, are also the actions a developer or an automation pipeline would perform. The result is a visibility paradox first articulated by the cloud forensics community more than a decade ago and still unresolved \cite{ruan2011cloudforensics, zawoad2013digital}: the logs are present, the events are captured, and yet the time from initial signal to investigator confidence is measured in hours or days. Industry threat reports and earlier survey work consistently identify reconstruction latency rather than detection coverage as the binding constraint in cloud incident response \cite{alharthi2024rlgnn}.

Two structural properties of cloud environments compound this latency. First, telemetry is intentionally granular. Cloud providers record each API as a separate event so that audit, billing, and compliance use cases can each project the data they need. A single attacker pulling three files generates a sequence of head, list, version, lock, and get calls that, viewed independently, look like routine reads. Second, the identity that performs those calls is itself layered. A long lived IAM user assumes a role through STS, the role issues a temporary session credential whose access key starts with the \texttt{ASIA} prefix, and the session calls services on behalf of the assumed principal. Reconstructing who actually acted requires joining across event sources that were never designed to be joined cheaply.

Deception offers a way out of this paradox by inverting the detection problem. Rather than attempting to recognize attacker behavior against an ever shifting baseline of legitimate behavior, a decoy resource generates signal only when it is touched, and that signal is by construction suspicious because no legitimate workflow has a reason to reach the decoy. The intellectual lineage is long, beginning with Spitzner's foundational treatment of honeypots \cite{spitzner2003honeypots} and extending through early work on decoy documents and bait injection \cite{bowen2009decoy, bowen2010botswindler} and through more recent cloud native measurements such as the honeybuckets study \cite{izhikevich2024honeybuckets}, which demonstrated that scanners do reliably target plausibly named S3 buckets and do engage with their contents. What the deception literature has shown is that the signal to noise advantage of a decoy is enormous; what it has historically not addressed is what to do once the decoy fires. Most deployments still hand the alert to a human analyst who must perform the same reconstruction the unanchored detection pipeline would have demanded.

\subsection{What Is Actually Hard About Connecting a Decoy to an Agent}
\label{sec:hard}
An obvious objection to the design we present is that agentic workflows are now widely understood and widely deployed, that language models have been applied to intrusion detection many times over, and that piping a decoy alert into such a workflow is therefore an integration exercise rather than a research contribution. We take that objection seriously and state directly where we believe it fails. Three properties of cloud telemetry turn the connection into a design problem with non obvious answers, and the framework is organized around them.

The first is that the unit of investigation is not the unit of logging. A decoy fires on an event, but no single event supports a forensic conclusion, and the natural grouping key is not available in the record. CloudTrail attributes an action to whichever credential performed it, which for a federated adversary is an ephemeral session credential rather than the durable principal an analyst needs in order to remediate. Grouping on the field the provider supplies fragments a single campaign across as many groups as the adversary has sessions; grouping too coarsely merges unrelated actors behind a shared network address. Section~\ref{sec:formal} defines the pivot tuple we group on and the aggregation operator it induces.

The second is that the agent's evidence horizon has to be bounded, and bounding it correctly is what makes the deception signal load bearing rather than decorative. An agent with unrestricted query access to control plane history will, given a suggestive session, retrieve context until its window fills, and the retrieved context will be overwhelmingly benign because that is what cloud accounts contain. The result is an agent that has inherited exactly the false positive profile and exactly the per invocation cost that deception was introduced to eliminate. The decoy is what makes a tight horizon defensible, because a tight horizon around a high confidence anchor discards much less than a tight horizon around a probabilistic one. Section~\ref{sec:cost} gives this argument its quantitative form and shows that per incident cost scales with decoy interaction volume rather than with total telemetry volume.

The third is the one we consider most consequential and which the prior literature on language models for log analysis has largely not confronted. Cloud telemetry is partly adversary authored. Object keys, request parameters, and user agent strings are attacker chosen values that the provider records verbatim and faithfully, and any pipeline that moves log fields into a model prompt has thereby constructed an indirect prompt injection channel \cite{greshake2023injection, owasp2025llmtop10}. Deception widens this rather than narrowing it, which cuts against the intuition that a decoy is a purely defensive surface. A decoy exists in order to be touched, and a decoy that permits writes is a resource on which an adversary can compose arbitrary strings in the expectation that a defender will read them. Section~\ref{sec:untrusted} develops the analysis and specifies the mitigation it requires. We are explicit that the prototype evaluated in this paper does not implement that mitigation, and we regard closing the gap as a precondition for any production deployment rather than as an optional refinement.

A fourth property is not specific to deception but conditions everything above. A language model asked to fill a report template will fill the template. Given ten standard fields and evidence for six, it will produce ten plausible values, and the four fabricated ones will be indistinguishable in tone from the six supported ones. The hallucination literature treats this as a primary failure mode in safety critical applications \cite{huang2023halsurvey}, and prior cloud forensics work applying language models has flagged exactly this risk \cite{alharthi2025llmforensics, alharthi2025ciaf}. Our response is to make the prompt itself carry the evidential boundary, which is what we mean by dynamic prompt generation.

\subsection{Contributions}
This paper describes Cloud Decoy AI Agent, a framework that pairs a deliberately attractive cloud decoy with a tool using language model agent and a two stage prompt construction discipline. We make five contributions.

We formalize decoy anchored session reconstruction, defining a pivot tuple over identity, credential, source address, and resource, an aggregation operator induced by that tuple, and a grounding invariant that any report generator over the resulting evidence package must satisfy (Section~\ref{sec:formal}).

We identify cloud telemetry as a partially adversary authored channel and show that deception anchored agentic pipelines are structurally exposed to indirect prompt injection through fields the provider records faithfully. We specify the quarantine discipline this requires and state that the prototype does not yet implement it (Sections~\ref{sec:untrusted} and~\ref{sec:futureinjection}).

We give explicit construction rules for the incident attack graph, including node and edge typing and the closed world condition under which no node exists without a supporting observed field, which was previously left implicit (Section~\ref{sec:graphrules}).

We document the deployment configuration, the model configuration, and the full scenario catalog at the level of detail that permits an independent reconstruction of the evaluation, and we report metrics under names that describe what they measure rather than implying a population level rate the design does not support (Sections~\ref{sec:impl} through~\ref{sec:results}).

We give an analytic cost model showing that per incident cost in a deception anchored design is linear in decoy interaction rate and independent of total account activity, which is the precise sense in which we claim deception changes the engineering economics of agentic investigation (Section~\ref{sec:cost}).

\begin{figure*}[t]
\centering
\includegraphics[width=0.85\textwidth]{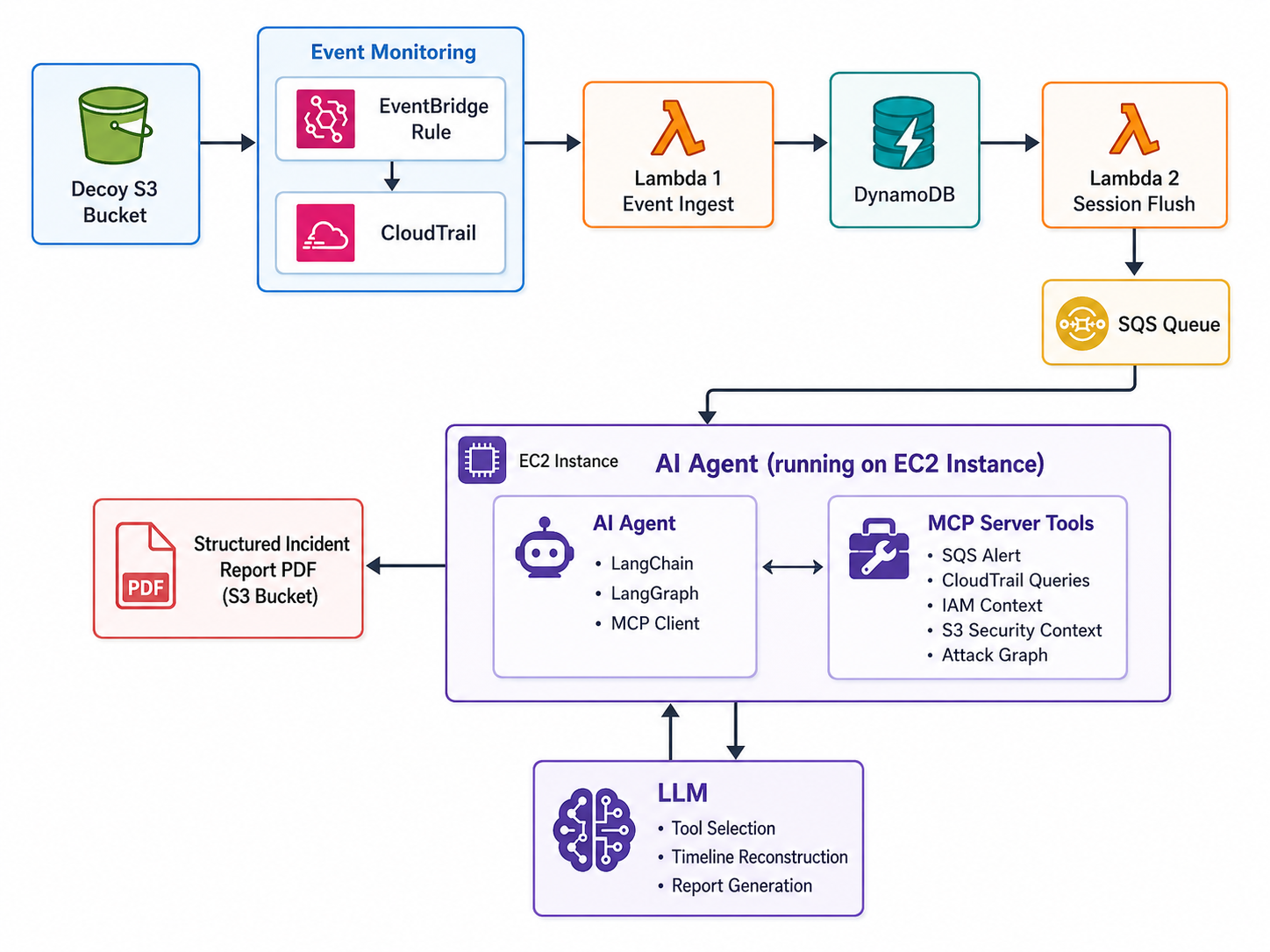}
\caption{End to end architecture of Cloud Decoy AI Agent. The decoy bucket receives suspicious interactions whose CloudTrail data events are routed through EventBridge into a stateful ingest function. Sessions are flushed through SQS to the agent, which calls MCP exposed tools for CloudTrail, IAM and STS, and S3 context. The LLM produces a grounded narrative that is rendered as the final structured report.}
\label{fig:arch}
\end{figure*}

% =====================================================================
\section{Background and Related Work}
% =====================================================================
The framework draws on three threads that have historically run in parallel: cloud forensics, deception based detection, and tool using language model agents. The decision to combine them rests on the observation that each of the three solves a problem that the other two have left open.

\subsection{Cloud Forensics and the Latency Problem}
Foundational work on cloud forensics laid out the canonical taxonomy of challenges \cite{ruan2011cloudforensics}, including distributed evidence, multi tenancy, opaque provider infrastructure, and the absence of physical custody, all of which constrain the depth at which a customer can ever investigate. Subsequent meta studies formalized the proactive forensics direction \cite{zawoad2013digital}, arguing that the practical fix is to instrument cloud workloads ahead of time and to design pipelines that assume the provider will not be a participant in investigation. More recent work has explored evidence accumulation, forensic readiness as a service, and structured ontology driven analysis, with the Cloud Investigation Automation Framework \cite{alharthi2025ciaf} as a representative modern instance that uses ontology based validation and language model reasoning to standardize the investigation pipeline, and with related work applying few shot classification over forensic log corpora and comparing it against gradient boosted baselines \cite{alharthi2025llmforensics}. Migration and readiness strategies that determine which evidence exists at all have been treated separately \cite{alharthi2023scms}. The NIST definition of cloud computing \cite{mell2011nistcloud} remains the operative reference for the service model boundaries that determine which evidence customers can collect. Across this body of work the unifying gap is consistent: the constraint is rarely the absence of logs and is almost always the cost of reconstructing chronology, identity, and intent fast enough to act.

\subsection{Deception and Decoys}
Spitzner's treatment of honeypots in network security \cite{spitzner2003honeypots} established the principle that low interaction decoys provide high fidelity signal because any contact is by construction suspicious. Subsequent work extended the principle from network probes to file level decoys, demonstrating with BotSwindler and the broader decoy document line of research that believable bait can be injected at the host level to detect masqueraders and crimeware \cite{bowen2009decoy, bowen2010botswindler}. The cloud era brought a new substrate. Storage buckets, IAM roles, secrets in parameter stores, and serverless functions can all be staged as decoys with negligible operating cost. The honeybucket measurement study \cite{izhikevich2024honeybuckets} provided large scale empirical evidence that S3 scanners actively engage with plausibly named buckets, including downloading and reading content. Recent containerized honeypot deployments \cite{srinivasa2023container} and comprehensive surveys of deception techniques \cite{javadpour2024deception} reinforce that the deception primitive is mature; what those efforts do not fully address is what happens between alert and investigator. Our work treats deception not as the destination but as the source of a downstream agentic workflow, and additionally treats the decoy as an adversary controlled write surface whose contents enter the analysis path.

\subsection{Language Model Agents and Cloud Security}
The agent paradigm relevant here is the one introduced by ReAct \cite{yao2023react}, in which a language model interleaves natural language reasoning steps with structured tool invocations and observations of tool output. The underlying capacity for multi step reasoning was made plausible by chain of thought prompting \cite{wei2022chainofthought}, and the engineering substrate for orchestrating such agents over heterogeneous backends has matured around frameworks like LangGraph \cite{langchain2024langgraph} and around standardized tool protocols such as the Model Context Protocol \cite{anthropic2024mcp}. Retrieval augmented generation \cite{lewis2020rag} sits adjacent to this line of work and motivates the broader principle that grounding model output in retrieved evidence improves factuality, a principle that survey work on hallucination \cite{huang2023halsurvey} treats as load bearing for safety critical applications. Recent work on provenance native answer traces goes further, decomposing transparency into answer correctness, citation fidelity against supporting evidence, and per document influence measured under leave one resource out intervention \cite{faizan2026provenai}, which is the direction the per claim citation extension of Section~\ref{sec:limitations} points toward.

Within cybersecurity the applications have grown rapidly. Recent surveys catalog language model use across cyber defense \cite{hassanin2024llmcyber}. LogPr\'ecis demonstrated that fine tuned language models can label Unix shell attack tactics directly from raw logs \cite{boffa2024logprecis}. LLMCloudHunter extracts cloud detection rules from open source intelligence \cite{schwartz2024llmcloudhunter}. Multi agent collaboration frameworks for incident response have been explored \cite{hays2024multiagentir}, as has end to end autonomous network incident response with language model agents \cite{li2025autoir}. The Model Context Protocol itself has attracted formal security analysis showing that tool exposure creates an exploitable capability surface \cite{radosevich2025mcpaudit}. Adjacent work explores reinforcement learning and graph neural network approaches to cloud incident response \cite{alharthi2024rlgnn} and zero trust policies learned for constrained response settings \cite{alharthi2024zerotrust}. Content based detection pipelines in adjacent domains illustrate the static and dynamic analysis split that our timeline classifier echoes \cite{omara2023phishing}.

\subsection{Positioning}
Table~\ref{tab:related} states the delta along the axes that separate this work from its neighbors. Two design choices distinguish Cloud Decoy AI Agent. The first is the use of a deception signal as the upstream gate that defines the agent's investigation universe rather than as an alert that terminates in a queue. The second is a two stage prompt assembly that constructs the report prompt from evidence the agent itself fetched, under a stated grounding invariant, rather than from a static template or from retrieval over a generic corpus. These choices interact. The decoy bounds what the agent can plausibly be asked about, and the dynamic prompt bounds the answer by what the agent actually saw. A third element, the recognition that log fields are partially adversary authored and that this makes the pipeline an injection surface, does not appear in any of the prior systems we surveyed, and we treat it as an analysis contribution rather than as an implemented one.

\begin{table*}[t]
\caption{Positioning against representative prior systems.}
\label{tab:related}
\centering
\footnotesize
\renewcommand{\arraystretch}{1.2}
\begin{tabular}{@{}lllll@{}}
\toprule
System & Trigger & Agentic tool use & Grounding discipline & Telemetry treated as untrusted \\
\midrule
LogPr\'ecis \cite{boffa2024logprecis}          & Log corpus (offline)      & No (fine tuned classifier) & Supervised labels          & No \\
LLMCloudHunter \cite{schwartz2024llmcloudhunter} & CTI corpus              & No                         & Rule extraction validation & No \\
CIAF \cite{alharthi2025ciaf}                    & Log corpus (offline)      & No                         & Ontology based validation  & No \\
Tabletop IR agents \cite{hays2024multiagentir}  & Scenario script           & Yes (multi agent)          & None stated                & No \\
Autonomous network IR \cite{li2025autoir}       & Network anomaly           & Yes                        & In context examples        & No \\
Containerized honeypot \cite{srinivasa2023container} & Decoy interaction    & No                         & Not applicable             & No \\
\midrule
\textbf{This work}                               & \textbf{Decoy interaction} & \textbf{Yes (MCP tools)}   & \textbf{Evidence bounded prompt (Inv.~1)} & \textbf{Identified, not yet mitigated} \\
\bottomrule
\end{tabular}
\end{table*}

% =====================================================================
\section{Threat Model and Design Goals}
\label{sec:threat}
% =====================================================================
The framework assumes a cloud tenant operating an account or organization on a major IaaS provider, with control plane logging enabled and with S3 data events captured by CloudTrail. The adversary is presumed to have obtained working credentials through any of the standard paths documented in the MITRE ATT\&CK enterprise cloud matrix \cite{strom2020attack, mitre2024cloud}, including phishing of a developer, theft of a long lived access key, exploitation of a SaaS to cloud trust relationship, or compromise of a CI/CD secret. Credential acquisition through human targeting remains the dominant initial access path, and prior survey work quantifies both the defense mechanisms available and the low level of employee awareness that makes the path reliable \cite{alharthi2021seipsurvey, alharthi2020sedefense, alharthi2020setaxonomy}. We do not assume the adversary lacks legitimate cloud knowledge; we assume only that they have no legitimate reason to interact with the decoy. The intrusion kill chain \cite{hutchins2011killchain} provides the broader conceptual frame for the stages that any such adversary moves through; our system aims to compress investigation latency within the stages from reconnaissance to actions on objectives.

\subsection{Telemetry as a Partially Adversary Authored Channel}
\label{sec:untrusted}
Beyond the conventional model we make one assumption that is unusual in this literature and that we believe should become standard for systems of this kind. We assume the adversary can author content that appears verbatim in the telemetry the system analyzes. Object keys are chosen by whoever writes or requests them. Request parameters, prefixes, delimiters, and user agent strings are client supplied. A provider that records these faithfully, as providers should, thereby carries adversary authored strings into every downstream consumer of the log. When the downstream consumer is a language model, the log is an untrusted input channel and the situation is structurally identical to indirect prompt injection through retrieved documents \cite{greshake2023injection, owasp2025llmtop10}.

Deception increases rather than decreases this exposure. A decoy exists to be touched, and a decoy that permits writes is a resource on which the adversary can compose arbitrary strings with the expectation that a defender will read them. An adversary who suspects instrumentation has an incentive to write object keys that read as instructions to an automated analyst. The concrete risks are that the generated report adopts adversary supplied framing, that the agent is induced to issue tool calls outside the intended investigation scope, and that fabricated indicators of compromise reach an analyst facing artifact under the authority of a system that is trusted precisely because its other outputs are well grounded.

The mitigation this requires is a partition of the evidence into provider derived and client supplied fields, with the latter length bounded, escaped, and delimited before reaching the model, never read by correlation logic in raw form, and surrounded by an instruction identifying the region as data rather than direction. Section~\ref{sec:formal} carries the partition into the formalization and the pivot tuple is drawn only from the provider derived side, which is the one place the prototype does respect the distinction. The remaining elements are specified in Section~\ref{sec:futureinjection} and are not implemented in the system evaluated here. We state this plainly rather than deferring it to a limitations paragraph because a reader deciding whether to deploy a design of this shape needs to know it at the point where the threat model is set. The general principle, that prompt construction is a place to apply deterministic validation rather than to trust model judgment, follows the argument made for ontology validated prompt interfaces in prior work \cite{alharthi2025promptshield}.

Out of scope are denial of service attacks against the cloud provider itself, supply chain compromise of the provider's control plane, compromise of the model provider, and adversaries who never touch any decoy surface. The last is a genuine and unavoidable limitation of any deception anchored design and Section~\ref{sec:limitations} returns to it.

\subsection{Design Goals}
From this threat model we derive six design goals. The first is signal purity. Every alert that enters the agent should be a high confidence indicator rather than a probabilistic anomaly. The second is chronological fidelity. The reconstructed timeline must preserve the order in which events occurred, the temporal gaps between them, and the distinction between successful and failed access, because forensic conclusions are sensitive to these features in ways that summaries are not. The third is identity attribution. The system must associate activity with the correct durable IAM principal, not merely with the ephemeral credential that performed the call, because remediation actions hang on that attribution. The fourth is evidential grounding. Every claim in the final report must trace to a specific observed artifact, since unsupported claims actively mislead responders. The fifth is injection resistance. Adversary authored content must be reportable without being executable as direction. The sixth is operational latency. A usable system must move from initial trigger to analyst ready report on a timescale comparable to a human first responder's time to context, which in practice we set at under ten minutes.

These goals trade against one another in characteristic ways. Aggressive evidence enrichment improves grounding but extends latency and widens the injection surface. Strict grounding rules suppress fluency in the generated narrative. Tight session windows preserve chronological fidelity for compact attacks but underperform when the adversary deliberately introduces delays. The architecture below makes the trade offs explicit, and the fifth goal is the one the current prototype meets least.

% =====================================================================
\section{Problem Formalization}
\label{sec:formal}
% =====================================================================
We state the reconstruction problem precisely, because the specificity of the definitions is what separates the design decisions from arbitrary engineering choices.

Let $e$ denote a CloudTrail data event. We project each event onto the fields that matter for correlation and analysis, writing
\[
e = \langle t,\; \iota,\; \kappa,\; \rho,\; \beta,\; a,\; o,\; \epsilon,\; u \rangle,
\]
where $t$ is the event time, $\iota$ the source address, $\kappa$ the access key identifier, $\rho$ the principal identifier as recorded, $\beta$ the target bucket, $a$ the event name, $o$ the object key when present, $\epsilon$ the error code when present, and $u$ the user agent string. We partition these fields into a provider derived set $T = \{t,\iota,\kappa,\rho,\beta,a,\epsilon\}$, which the provider computes from authenticated request context, and a client supplied set $A = \{o,u\}$, which the provider records as the client presented it. This partition is the formal statement of the assumption in Section~\ref{sec:untrusted} and it propagates through the rest of the design.

\paragraph*{Pivot tuple} Define the pivot projection $\pi(e) = \langle \iota, \kappa, \rho, \beta \rangle$. The pivot is drawn entirely from $T$, so an adversary cannot influence how their own activity is partitioned by choosing what to write. Note that $\rho$ as recorded is the acting credential's principal, which for a federated adversary is the assumed role session rather than the durable identity; resolving $\rho$ to a durable principal is an enrichment step, not a parsing step, and Section~\ref{sec:tools} describes the tool that performs it.

\paragraph*{Session} Given a decoy event stream $E$ ordered by $t$ and an idle threshold $\tau$, a session $S \subseteq E$ is a maximal subset such that all events share a pivot, $\pi(e) = \pi(e')$ for all $e,e' \in S$, and consecutive events in $S$ are separated by no more than $\tau$. Writing $S = \langle e_1,\dots,e_n\rangle$ in time order, $t_{i+1} - t_i \le \tau$ for all $i < n$, and no event outside $S$ with the same pivot falls within $\tau$ of an endpoint. The prototype uses $\tau = 15$ minutes.

\paragraph*{Campaign} A campaign $C$ is a set of sessions that a competent analyst would attribute to a single adversarial episode. Sessions and campaigns coincide when the adversary acts compactly and diverge when the adversary introduces gaps exceeding $\tau$. The delayed access failure reported in Section~\ref{sec:delayed} is exactly the case $|C| > 1$ with the system reporting $|C|$ separate incidents.

\paragraph*{Evidence package} Enrichment maps a session to an evidence package $P = \langle S, I, X, G \rangle$, where $I$ is the resolved identity context including role chain, $X$ the resource security context, and $G$ the constructed attack graph.

\paragraph*{Grounding invariant} Let $R$ be the generated report and let $\mathrm{claims}(R)$ be the set of factual assertions it contains, where a factual assertion is any statement naming an identifier, a timestamp, an object, an event, a count, or a byte quantity. We require:

\begin{quote}
\textbf{Invariant 1 (Evidence boundedness).} For every $c \in \mathrm{claims}(R)$ there exists $p \in P$ such that $c$ is entailed by $p$, and no $c$ asserts a value for a field absent from $P$.
\end{quote}

Invariant 1 is not enforceable by inspection of the model, so the design enforces it structurally. The prompt is constructed to contain exactly $P$ and nothing else, and missing fields are omitted rather than represented by placeholders. Section~\ref{sec:results} reports the empirical violation rate over the scenario suite, which is the measurement that most directly bears on the claim made for dynamic prompt generation.

% =====================================================================
\section{System Architecture}
% =====================================================================
The system is organized as a six layer pipeline. From the decoy at the edge of the cloud environment to the structured report at the analyst's desk, each layer is responsible for one reduction of ambiguity, and the interfaces between layers are deliberately narrow so that any component can be replaced or extended without disturbing the others. Figure~\ref{fig:arch} shows the full data path.

\subsection{Decoy Resource Design}
The decoy is an Amazon S3 bucket named to evoke a sensitive enterprise asset. In the validation deployment we used \texttt{corp\_finance\_archive\_2023} and populated it with realistically named but fabricated objects such as \texttt{api\_keys.txt}, \texttt{database\_backup.sql}, and a small set of finance themed files. The names matter. The honeybucket measurement study \cite{izhikevich2024honeybuckets} found measurable selection bias in scanner targeting based on bucket plausibility, and the honeyfile literature \cite{whitham2020honeyfilesurvey} has long argued that the camouflage of a decoy is determined by how naturally its filenames blend into the environment. The bucket has no bucket policy and all public access block settings enabled, which forces any access to route through identity based authorization rather than through anonymous public reads, and which ensures that the IAM principal and access key are recorded on every event.

A key property of the decoy is that it must be reachable through the tenant's normal IAM trust graph, because the threat model concerns adversaries who have obtained working credentials. The decoy therefore lives in the same account and region as production resources and is intentionally not isolated by a hard network boundary. Its security comes not from inaccessibility but from the fact that no legitimate workflow has a reason to touch it. How well that property survives an adversary who is actively looking for instrumentation is untested here and is treated in Section~\ref{sec:limitations}.

\subsection{Event Detection and Routing}
S3 data events for the decoy bucket are captured by CloudTrail and forwarded through an EventBridge rule to a Lambda function that performs initial normalization. The normalized record is keyed by the pivot tuple $\pi(e)$ and written to a DynamoDB table whose primary key is that tuple plus a session identifier. A second Lambda runs on a scheduled timer and flushes sessions that have been idle for longer than $\tau$. A session flush emits a single structured message onto an SQS queue that the agent consumes.

This design choice deserves comment. We considered three alternatives. A streaming approach that sends every event directly to the agent would have minimized latency for compact attacks but would have multiplied agent invocations on bursty scripted access, and would have lost the chronological signal that comes from seeing a sequence of related events as one unit. A batch approach that runs the agent on a fixed schedule would have been simpler but would have introduced a worst case delay equal to the batch interval. The hybrid design with idle based session flushing preserves both compactness for the common case and bounded latency for the worst case, at the cost of making the flush window the dominant term in end to end latency, which Section~\ref{sec:latency} quantifies.

\begin{figure*}[t]
\centering
\includegraphics[width=0.85\textwidth]{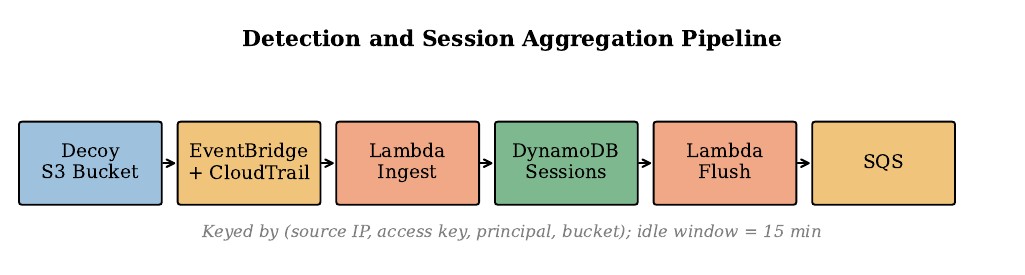}
\caption{Detection and aggregation pipeline. Raw CloudTrail events on the decoy are routed through EventBridge, normalized by an ingest Lambda, accumulated in DynamoDB keyed by the pivot tuple, and flushed as a session message onto SQS for the agent to consume. Idle based flushing trades fixed latency for chronological grouping.}
\label{fig:pipeline}
\end{figure*}

\subsection{Session Aggregation}
Cloud logs are intentionally granular, and the granularity is the enemy of reconstruction. A single attacker action such as enumerating a bucket and pulling two files can produce eight or more CloudTrail records once head, list, version, lock, and get calls are accounted for. Treating each record as a separate incident produces eight alerts and zero context. The aggregation operator of Section~\ref{sec:formal} groups records by the pivot and bounds each group by the idle threshold. Within a session, events are sorted by event time and the sequence of event names is preserved as an ordered list.

The choice of pivot fields is informed by what subsequent enrichment will need rather than by what the raw event already contains, and it is deliberately restricted to the provider derived partition $T$. CloudTrail records contain dozens of fields, but only a small subset are useful for cross event correlation, and none of the useful ones are client supplied. Grouping on a client supplied field such as the user agent would hand the adversary direct control over how their own activity is partitioned.

\subsection{Agent Runtime and MCP Tools}
\label{sec:tools}
The agent runs on a small EC2 instance that consumes session messages from SQS. The implementation uses LangChain for primitives and LangGraph \cite{langchain2024langgraph} for the stateful workflow, with the Model Context Protocol \cite{anthropic2024mcp} as the tool interface. Each MCP tool exposes one capability that the agent can invoke without knowing how the capability is implemented underneath. Table~\ref{tab:tools} gives the interface of each tool and the IAM permissions it requires, which together constitute the capability surface that a security audit of the deployment would review \cite{radosevich2025mcpaudit}. Each tool runs under a distinct execution role holding only the permissions listed, so an agent induced to make an unintended call still cannot reach a capability the tool does not hold.

\begin{table*}[t]
\caption{MCP tool interfaces and the capability surface they expose.}
\label{tab:tools}
\centering
\scriptsize
\renewcommand{\arraystretch}{1.25}
\begin{tabular}{@{}p{0.16\textwidth}p{0.22\textwidth}p{0.28\textwidth}p{0.26\textwidth}@{}}
\toprule
Tool & Input & Output & IAM permissions required \\
\midrule
\texttt{cloudtrail\_history} & principal, source address, or time range & ordered event records projected onto $T \cup A$ & \texttt{cloudtrail:\linebreak[0]LookupEvents} \\
\texttt{identity\_context}   & principal identifier or access key & role chain, credential type, temporary session flag, durable principal & \texttt{iam:GetUser}\newline \texttt{iam:GetRole}\newline \texttt{sts:GetCaller\-Identity} \\
\texttt{s3\_security\_context} & bucket name & bucket policy, public access block state, versioning, object lock & \texttt{s3:GetBucket\-Policy}\newline \texttt{s3:GetBucket\-PublicAccess\-Block}\newline \texttt{s3:GetBucket\-Versioning} \\
\texttt{attack\_graph}       & evidence package & typed node and edge set (Sec.~\ref{sec:graphrules}) & none, operates locally \\
\bottomrule
\end{tabular}
\end{table*}

The use of MCP rather than provider specific function calling matters for three reasons. Tools become portable across model vendors, since any MCP compliant client can speak to the same server. New tools can be added without retraining or reconfiguring the model. And the formal security framework that has begun to form around MCP \cite{radosevich2025mcpaudit} gives a clear vocabulary for reasoning about which capabilities the agent should and should not have.

\subsection{Reasoning and Report Generation}
\label{sec:models}
The reasoning layer uses one model as the controller and a separate model invocation for the final report generation. The split mirrors the cost and quality trade off that practical agent systems consistently make. A fast model orchestrates tool calls and routes the workflow, and a higher quality model produces the final analyst facing prose. Model identifiers and decoding parameters are given in Table~\ref{tab:config}. The output is rendered as a structured report containing an executive summary, an ordered event timeline, identity analysis, S3 access analysis, a reconstructed attack pattern, indicators of compromise, and recommendations.

% =====================================================================
\section{The AI Agent Workflow}
\begin{figure}[t]
\centering
\includegraphics[width=0.95\columnwidth]{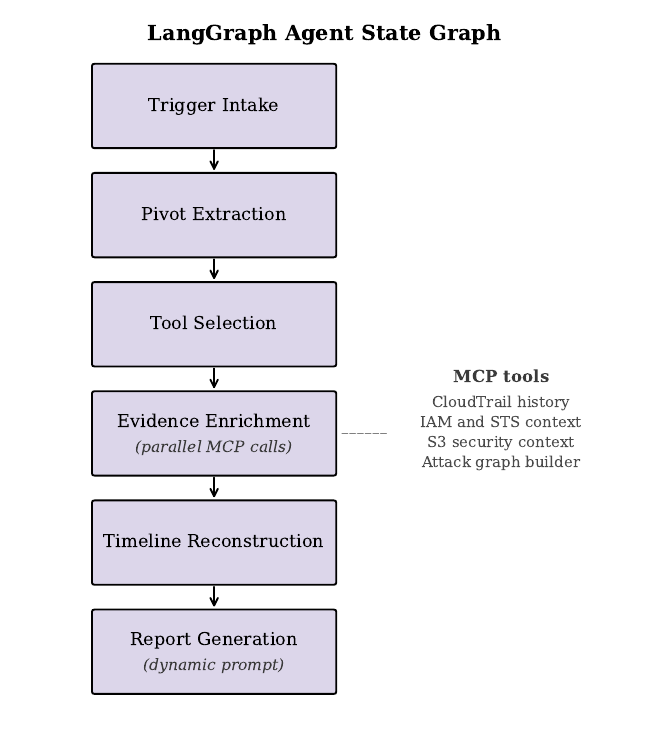}
\caption{Agent state graph. Trigger intake parses the SQS payload. Pivot extraction identifies which enrichment tools are needed. Tool selection issues parallel or sequential MCP calls. Evidence enrichment merges results. Timeline reconstruction orders the events. Report generation assembles the dynamic prompt and produces the final report.}
\label{fig:agentgraph}
\end{figure}

% =====================================================================
The agent receives a session message and walks through six stages bound to the LangGraph state graph in Figure~\ref{fig:agentgraph}.

\subsection{Trigger Intake and Pivot Extraction}
The first stage validates the session message and rejects malformed entries rather than repairing them. The second stage extracts the pivots that the rest of the workflow will use, together with the observed object keys, the sequence of event names, the timestamps, and any error codes. The pivot dictionary is the only piece of state that survives across all subsequent stages, and its cleanliness determines the cleanliness of the eventual report. A failure here propagates downstream, so the stage is conservative. It prefers to mark a field as missing rather than to guess, which is the operational form of Invariant 1 at the input boundary.

\subsection{Tool Selection and Evidence Enrichment}
With the pivot dictionary in hand the agent decides which MCP tools to call. The decision is not hard coded; the agent reasons over the available pivots and chooses the tools that the available evidence justifies. In practice \texttt{cloudtrail\_history} is almost always invoked because the source address and access key are almost always present. \texttt{identity\_context} is invoked when the principal identifier resolves or when the access key carries the temporary session prefix \texttt{ASIA}. \texttt{s3\_security\_context} is invoked when the bucket name resolves and when the analysis depends on whether the bucket was publicly accessible. Tool calls run in parallel when their inputs are independent, which is the common case, and the merged result is stored in the evidence package.

\subsection{Timeline Reconstruction}
Timeline reconstruction orders the events strictly by event time, classifies each as enumeration, retrieval, denied, or destructive, preserves temporal gaps as explicit fields rather than collapsing them, and assembles a chronological skeleton that the report generator renders in prose. The classifier is deliberately conservative. When an event does not unambiguously match one of the four classes it is preserved with its raw event name rather than reclassified, on the reasoning that an unclassified entry is a smaller error than a confidently misclassified one and is visible to the analyst as such.

\subsection{Dynamic Prompt Generation}
The technical contribution of the workflow is the way the report generator's prompt is built. A static system prompt establishes role, output structure, and constraints. It states that the model must use only the supplied evidence, that it must preserve temporal ordering, that it must distinguish successful from failed access, that it must list each accessed object explicitly when retrieval occurred, and that it must not introduce indicators of compromise that did not appear in the evidence. The static prompt is identical across every invocation of the system.

\begin{figure}[t]
\centering
\includegraphics[width=0.95\columnwidth]{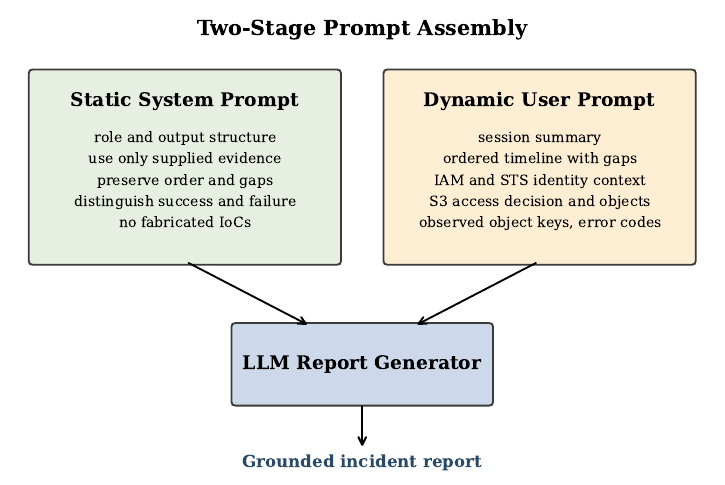}
\caption{Dynamic prompt assembly. The static system prompt defines the reasoning framework and constraints. The user prompt is generated per session from the agent's actual evidence package. Missing fields stay missing rather than being filled with placeholders, which removes a major class of hallucination triggers.}
\label{fig:prompt}
\end{figure}

The user prompt is generated fresh for every session from the evidence package the agent itself collected. It contains the session summary, the ordered event timeline with explicit gaps preserved, the CloudTrail artifacts including event identifiers and bytes out values, the IAM and STS identity details with the role chain when relevant, the S3 access decision observations including bucket policy and public access block state, the list of object keys, and any error codes that occurred. Crucially, the user prompt does not include any field that the agent did not actually observe. Missing pivots remain absent rather than being filled with placeholder text, because a placeholder is an invitation to invent a value that satisfies it. Figure~\ref{fig:prompt} illustrates the assembly.

This two stage discipline is consistent with what the hallucination literature has converged on. Separating the reasoning framework from the evidence context, and grounding generated claims in retrieved or observed material, materially reduces fabrication \cite{huang2023halsurvey, lewis2020rag}. It also makes the system extensible, since a new evidence source extends the user prompt without rewriting the system prompt and a new reporting constraint extends the system prompt without disturbing the evidence pipeline. We note the limitation this evaluation carries: the design is not compared against a static template condition, so the measurements in Section~\ref{sec:results} establish that the system as built satisfies Invariant 1 across the suite but do not isolate how much of that is attributable to prompt assembly rather than to the enrichment that precedes it. Section~\ref{sec:missing} names the comparison that would settle it.

\subsection{Attack Graph Construction}
\label{sec:graphrules}
The final stage emits a graph representation of the incident. Because the construction rules were previously left implicit, we state them.

The graph $G = (V,E)$ is typed. Node types are $\mathsf{Principal}$, $\mathsf{Credential}$, $\mathsf{Address}$, $\mathsf{Resource}$, and $\mathsf{Object}$. Edge types are $\mathsf{authenticated\_as}$ from $\mathsf{Address}$ to $\mathsf{Principal}$, $\mathsf{assumed}$ from $\mathsf{Principal}$ to $\mathsf{Principal}$ along a resolved role chain, $\mathsf{used}$ from $\mathsf{Principal}$ to $\mathsf{Credential}$, $\mathsf{acted\_on}$ from $\mathsf{Credential}$ to $\mathsf{Resource}$ labeled with the event name and outcome, and $\mathsf{touched}$ from $\mathsf{Resource}$ to $\mathsf{Object}$.

Construction proceeds by three rules. Rule one instantiates one node per distinct identifier appearing in the evidence package. Rule two instantiates one $\mathsf{acted\_on}$ edge per event in the reconstructed timeline, carrying the event time, the event name, and the success or failure outcome, so that edge multiplicity preserves repeated access rather than collapsing it. Rule three instantiates $\mathsf{assumed}$ edges only where the identity tool returned an explicit role chain, never by inference from naming conventions.

\begin{figure}[t]
\centering
\includegraphics[width=0.95\columnwidth]{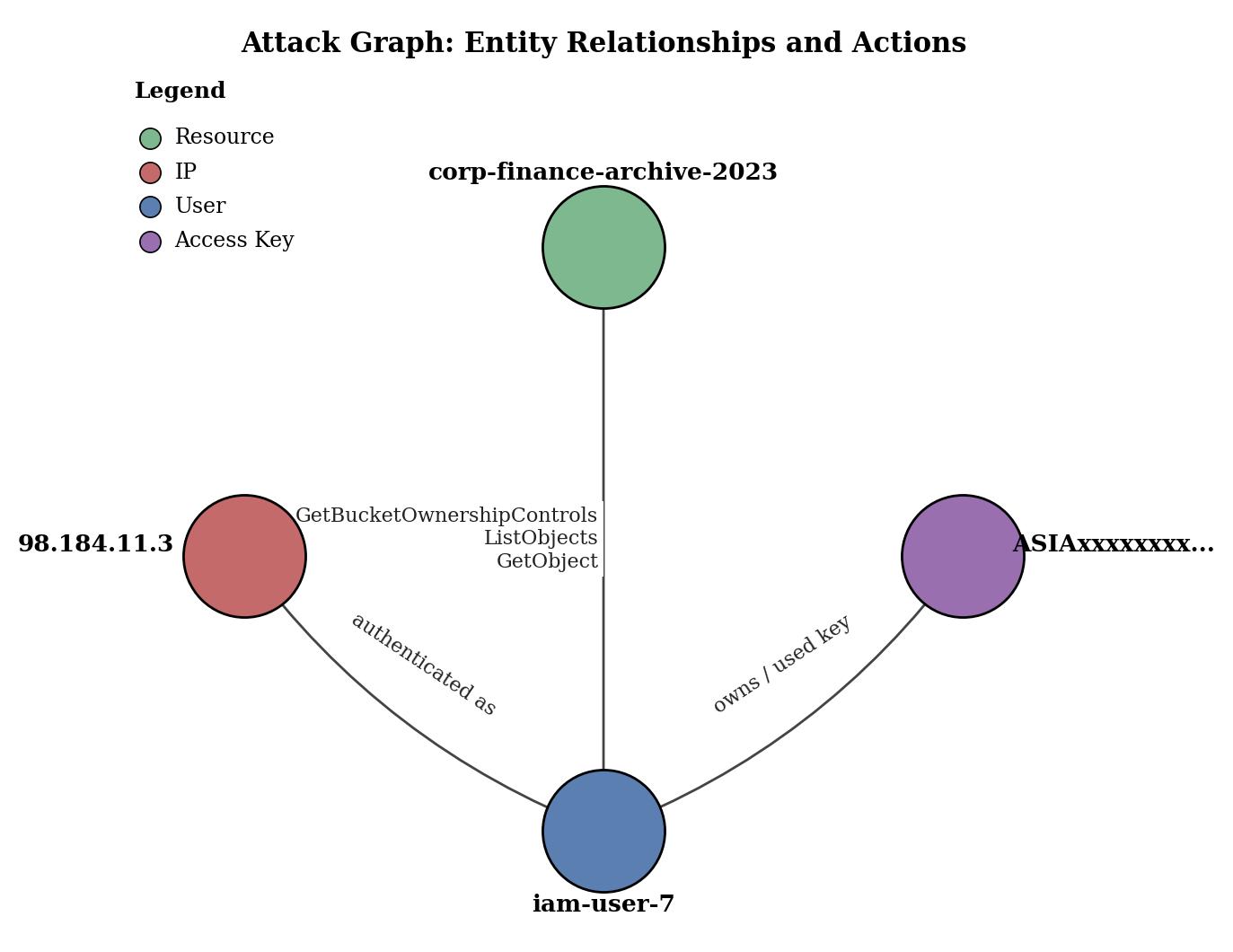}
\caption{Reconstructed attack graph for a representative validation incident on the decoy bucket. The IAM principal authenticated from a single source address, used the indicated access key, and performed enumeration followed by retrieval of \texttt{api\_keys.txt}. The graph is generated by the agent and embedded in the final report.}
\label{fig:attackgraph}
\end{figure}

Because the graph is constructed exclusively from fields present in the evidence package, it satisfies a closed world condition: no node and no edge exists without a supporting observed field. This makes the graph a structural witness for Invariant 1 as well as a visualization, and it is the reason the same structure serves both as the representation the agent pivots over during enrichment and as the diagram embedded in the final report. Graph based representations of cyber incidents have a long lineage \cite{sheyner2002attackgraph}; what we add is the explicit statement of the closure condition that ties the graph to observed evidence rather than to inferred structure. Figure~\ref{fig:attackgraph} shows the reconstruction for a representative incident from the validation set.

% =====================================================================
\section{Experimental Methodology}
\label{sec:impl}
% =====================================================================

\subsection{Environment and Configuration}
We evaluated the framework on AWS in a dedicated account separated from any production workload. The decoy bucket and the supporting Lambda, EventBridge, DynamoDB, and SQS components were deployed in that account, and the agent ran on a small EC2 instance in the same account and region. Table~\ref{tab:config} records the configuration. We report model identifiers and decoding parameters explicitly, because the central claim of the paper concerns a prompt construction discipline and a reader cannot judge whether that discipline is model agnostic without knowing which model produced the reported behavior. We did not run the suite under a second model family, so the question of portability across vendors remains open and we make no claim about it.

\begin{table}[t]
\caption{Deployment and model configuration.}
\label{tab:config}
\centering
\renewcommand{\arraystretch}{1.15}
\begin{tabular}{@{}p{0.40\columnwidth}p{0.50\columnwidth}@{}}
\toprule
Parameter & Value \\
\midrule
Cloud provider          & AWS, single account, single region \\
Account topology        & dedicated, no production workload \\
Telemetry source        & CloudTrail S3 data events on decoy \\
Routing                 & EventBridge rule to ingest Lambda \\
Session store           & DynamoDB, keyed by pivot tuple \\
Idle threshold $\tau$   & 15 min \\
Flush transport         & SQS, one message per session \\
Agent host              & EC2, general purpose small instance \\
Orchestration           & LangChain primitives, LangGraph state machine \\
Tool interface          & Model Context Protocol, four tools \\
Controller model        & \controllermodel \\
Report model            & \reportmodel \\
Report output           & structured PDF with embedded attack graph \\
\bottomrule
\end{tabular}
\end{table}

\subsection{Scenario Catalog}
\label{sec:catalog}
Table~\ref{tab:catalog} documents the ten scenario suite. Each row states the scenario identifier, the behavioral family, the API behavior the driver script executes, and the CloudTrail artifact the scenario is designed to produce. The reconstruction requirement is uniform across scenarios and is stated once below rather than repeated per row.

Scenario families were chosen to span the behaviors we considered representative of decoy triggered cloud attacks: enumeration without retrieval, enumeration followed by retrieval of a single object, multi object retrieval after enumeration, an unauthorized access attempt resulting in an \texttt{AccessDenied} error, a destructive sequence in which retrieval is followed by deletion, a delayed sequence in which enumeration is separated from retrieval by a twenty minute gap, and script driven variants of four of these. Scripted variants were included because programmatic access produces tighter temporal clustering and a different distribution of user agent strings than interactive access, and a system that succeeds on interactive cases without handling scripted ones would not be operationally useful.

We acknowledge directly that a suite the authors constructed samples what we considered representative rather than what a community consensus would consider representative. No external taxonomy of decoy triggered cloud incidents exists against which to calibrate such a suite, and constructing one is an open problem for the deception community. Section~\ref{sec:validity} treats this as the principal construct validity threat.

\begin{table*}[t]
\caption{Scenario catalog. Ground truth is the executed action sequence recorded by the driver script rather than the captured CloudTrail record, so any divergence between executed action and captured event is itself observable.}
\label{tab:catalog}
\centering
\footnotesize
\renewcommand{\arraystretch}{1.2}
\begin{tabular}{@{}p{0.05\textwidth}p{0.20\textwidth}p{0.35\textwidth}p{0.31\textwidth}@{}}
\toprule
ID & Family & Executed API behavior & Expected CloudTrail artifact \\
\midrule
S1  & Enumeration only            & repeated bucket listing, no object access & list events only, no retrieval \\
S2  & Enumeration and retrieval   & listing followed by head and get on one object & list then get, with bytes out recorded \\
S3  & Multi object retrieval      & listing followed by get on several objects & one get event per object \\
S4  & Denied access               & get attempted under an unauthorized principal & \texttt{AccessDenied} error code \\
S5  & Destructive                 & retrieval followed by object deletion & get then delete \\
S6  & Delayed access              & enumeration, twenty minute gap, then retrieval & activity spanning two flush windows \\
S7  & Enumeration, scripted       & as S1 through the SDK & list events, SDK user agent \\
S8  & Enumeration and retrieval, scripted & as S2 through the SDK & list then get, SDK user agent \\
S9  & Multi object, scripted      & as S3 through the SDK & get events, SDK user agent \\
S10 & Denied access, scripted     & as S4 through the SDK & \texttt{AccessDenied}, SDK user agent \\
\bottomrule
\end{tabular}
\end{table*}

\subsection{Scoring Procedure}
\label{sec:scoring}
For each scenario we defined an objective, the simulated API behavior, the expected CloudTrail artifact, and the reconstruction requirement. The reconstruction requirement specified the minimum facts the final report had to contain: the correct classification of each event, the correct order, the correct attribution of the acting identity, and the absence of assertions unsupported by the evidence package. A scenario scored $1$ when the reconstruction met every requirement and $0.5$ when it met most but missed a non trivial element. We did not award fractional scores below half.

Scoring was performed by manual review against ground truth, with ground truth defined by the driver script rather than by the CloudTrail records, so that any gap between the executed action and the captured event would itself be observable. We state without qualification that the review was carried out by the authors and not by independent assessors, and that this is a material weakness. Section~\ref{sec:validity} treats it as such, and Appendix~\ref{app:rubric} reproduces the criteria applied so that the judgments are at least auditable in form even though they are not independent.

\subsection{Metrics}
\label{sec:metrics}
An earlier version of this work reported these measurements under names implying event level or population level rates. That presentation was not defensible, since the underlying counts are over scenarios rather than over events or over a defined population, and we have renamed the metrics to describe what they measure.

\textit{Complete event coverage} counts scenarios in which every ground truth API call appears in the reconstructed timeline. \textit{Correct ordering} counts scenarios in which every consecutive pair of ground truth events appears in the correct relative order. \textit{Correct attribution} counts scenarios in which the report names the correct IAM principal, the correct access key, and the correct role chain where one exists. \textit{Complete object naming} counts scenarios in which every accessed object key is named explicitly, over those scenarios in which retrieval occurred. \textit{Unsupported assertions} counts reports containing at least one factual assertion that cannot be traced to a field in the evidence package, that is, an observed violation of Invariant 1. \textit{Investigation latency} is wall clock time from the first CloudTrail event for a scenario to the moment the final report is written to its output bucket.

We deliberately do not report precision against a negative class. There is no meaningful population of true negatives in a decoy anchored design, since the decoy defines the positive set by construction, and a precision figure computed over an undefined negative class would be uninterpretable. Nor do we report interval estimates or significance tests. A suite of ten scenarios scored by the authors does not support them, and attaching them would lend the numbers an authority the design of the evaluation does not earn.

% =====================================================================
\section{Evaluation Results}
\label{sec:results}
% =====================================================================
The system reconstructed nine of the ten scenarios completely and one partially, for a suite score of $9.5/10$. Table~\ref{tab:scenarios} reports the per scenario outcome and Table~\ref{tab:metrics} the aggregate counts. The single partial result was the delayed access scenario, which we discuss separately because the failure mode is informative.

\begin{table}[t]
\caption{Per scenario reconstruction outcomes.}
\label{tab:scenarios}
\centering
\renewcommand{\arraystretch}{1.15}
\begin{tabular}{@{}p{0.62\columnwidth}c@{}}
\toprule
Scenario & Score \\
\midrule
S1 \; Enumeration only access & 1.0 \\
S2 \; Enumeration followed by retrieval & 1.0 \\
S3 \; Multi object retrieval & 1.0 \\
S4 \; \texttt{AccessDenied} attempt & 1.0 \\
S5 \; Retrieval followed by deletion & 1.0 \\
S6 \; Delayed access with twenty minute gap & 0.5 \\
S7 \; Enumeration only, script driven & 1.0 \\
S8 \; Enumeration plus retrieval, script driven & 1.0 \\
S9 \; Multi object retrieval, script driven & 1.0 \\
S10 \, \texttt{AccessDenied} attempt, script driven & 1.0 \\
\midrule
\textbf{Total} & \textbf{9.5/10} \\
\bottomrule
\end{tabular}
\end{table}

\begin{table}[t]
\caption{Aggregate outcomes across the ten scenario suite. Counts are over scenarios, not over events.}
\label{tab:metrics}
\centering
\renewcommand{\arraystretch}{1.15}
\begin{tabular}{@{}p{0.60\columnwidth}c@{}}
\toprule
Measure & Result \\
\midrule
Scenarios with complete event coverage & 9 of 10 \\
Scenarios with fully correct ordering & 9 of 10 \\
Scenarios with correct attribution & 9 of 10 \\
Scenarios with complete object naming & 6 of 6 \\
Reports containing an unsupported assertion & 0 of 10 \\
Investigation latency, trigger to report & 4 to 5 min \\
\bottomrule
\end{tabular}
\end{table}

Complete event coverage held in nine scenarios. The single miss was an enumeration event in the delayed access scenario that fell on the far side of the session flush window and was therefore assigned to a separate session that the agent processed as unrelated activity. Ordering was correct in the same nine scenarios, limited by the same cause, since the twenty minute gap pushed part of the activity into a downstream session. Attribution was correct in nine scenarios; the agent identified the IAM principal, the access key, and the role chain in every case except the delayed one, where the second session reconstruction lost the cross session link to the original principal context. Object naming was complete in all six scenarios in which retrieval was the operation of record, meaning every object key the driver script touched was named explicitly in the report.

No report in the suite contained an assertion the authors could not trace to a field in the evidence package. This is the measurement most directly relevant to the claim made for dynamic prompt generation, and it is also the measurement whose interpretation is most constrained by the absence of a comparison condition. It establishes that the system as built satisfied Invariant 1 across this suite. It does not establish that the prompt assembly discipline is what produced that result, and Section~\ref{sec:missing} names the comparison that would.

Figure~\ref{fig:latency} decomposes the latency, which is the right view for engineers reasoning about where to invest in further optimization.

\subsection{Latency}
\label{sec:latency}
\begin{figure}[t]
\centering
\includegraphics[width=0.95\columnwidth]{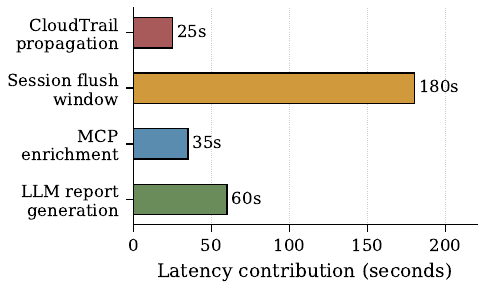}
\caption{Latency decomposition from first CloudTrail event to delivered report. The session flush window dominates and is configurable; CloudTrail propagation and MCP enrichment together contribute approximately a minute; report generation contributes the remainder.}
\label{fig:latency}
\end{figure}

End to end latency was four to five minutes from the first CloudTrail event to the delivered report. Figure~\ref{fig:latency} decomposes the interval. CloudTrail propagation contributed roughly twenty five seconds, the session flush window one hundred eighty seconds, MCP enrichment thirty five seconds, and report generation sixty seconds.

The composition matters more than the total. The flush window dominates and is a configuration parameter rather than a property of the system, so the headline figure is a statement about a parameter choice. Reducing $\tau$ would reduce latency proportionally and would shift the delayed access failure mode to shorter gaps. The remaining terms, propagation and inference, are the ones that bound how fast any design of this shape can be, and together they account for roughly two minutes. We report the total as a range across runs in a single region under a single configuration, and it should be read as an operational order of magnitude rather than as a figure that transfers unchanged to other regions, account structures, or model providers.

\subsection{Analysis of the Delayed Access Failure}
\label{sec:delayed}
The partial result on the delayed access scenario is worth examining because it isolates a genuine architectural trade off rather than an implementation defect. The session flush window was set to fifteen minutes, which is short enough to keep latency bounded for the common case but shorter than the twenty minute gap the scenario was designed to expose. When the gap exceeded the window, the enumeration session flushed and was processed as its own incident, and the retrieval session that followed was treated as a separate event. The agent reconstructed each of the two sub incidents correctly and identified the principal in both, but did not link them as parts of one campaign, which is what the reconstruction requirement asked for.

Extending the window is not the fix. It would move the failure mode to longer gap attacks while inflating latency for every compact one. The fix is a cross session correlation step above the flush layer that rejoins sessions whose pivots match within a longer horizon and whose combined sequence is consistent with a single campaign. Section~\ref{sec:futurecorr} states that procedure as an algorithm. It is not implemented in the system evaluated here, and the delayed access result therefore stands as an open failure rather than as one this paper closes.

\subsection{Sensitivity to Decoy Realism}
We performed a small qualitative check on how decoy realism affects the upstream signal. Buckets with plausibly named contents received engagement consistent with what the broader honeybucket measurement reported \cite{izhikevich2024honeybuckets}; buckets with placeholder names and obviously fake contents received markedly less, which matches the older honeyfile literature on camouflage \cite{whitham2020honeyfilesurvey}. We did not attempt to reproduce the large scale honeybucket measurement. We rely on that prior work and report only that decoy realism sits in the operative path for this kind of system and should be treated as a design parameter rather than as an afterthought. This observation is qualitative and we do not draw a quantitative claim from it.

\subsection{Threats to Validity}
\label{sec:validity}
Construct validity is bounded in two ways. Ground truth is defined by the driver scripts rather than by an external annotator's reading of each incident, which is standard for controlled evaluation but admits the possibility that the evaluation rewards behaviors aligned with our reconstruction model rather than with what an independent analyst would judge correct. More seriously, the scoring was performed by the authors, who also built the system. We do not regard the rubric in Appendix~\ref{app:rubric} as a substitute for independent assessment; it makes the criteria auditable but it does not remove the incentive. A future evaluation should present reports in randomized order with condition identifying material removed, to assessors outside the development team, with inter rater agreement reported.

Internal validity is bounded by the fixed fifteen minute flush window, which determined where the delayed scenario lost continuity. The failure mode would shift rather than disappear under a different window, and a sweep across window sizes would tighten the claim beyond what a single configuration supports.

External validity rests on three assumptions we cannot discharge here. The first is that scanner behavior against plausibly named cloud resources generalizes from the honeybucket measurement \cite{izhikevich2024honeybuckets} to the specific naming and content choices used in this deployment. The second is that a single region's CloudTrail propagation profile is representative, so the latency figures should be read as an order of magnitude rather than as a portable constant. The third is that results obtained with the models in Table~\ref{tab:config} transfer to other model families, which we did not test.

Statistical conclusion validity is minimal by construction. Ten scenarios scored by the authors support descriptive reporting and nothing stronger. We make no significance claims and we draw no conclusions from differences that a suite of this size cannot resolve.

% =====================================================================
\section{Cost and Scaling Analysis}
\label{sec:cost}
% =====================================================================
A question this class of system has to answer is what an investigation costs and how that cost behaves as the environment grows. We answer analytically rather than by quoting a figure from one week of one deployment, because a model a reader can instantiate against their own rates and their own traffic is more useful than a number that reflects ours.

Let $\lambda$ be the rate of decoy interactions per day and $\sigma$ the mean number of sessions per interaction burst, so that the daily agent invocation count is $\lambda\sigma$. Per incident cost decomposes into infrastructure and inference terms:
\begin{equation*}
\begin{split}
c_{\text{inc}} \;=\;
  & \underbrace{c_{\text{ct}} n_e + c_{\lambda} d + c_{\text{ddb}} n_w + c_{\text{sqs}}}_{\text{infrastructure}} \\[2pt]
  & +\; \underbrace{\sum_{m} \left( p^{\text{in}}_m T^{\text{in}}_m + p^{\text{out}}_m T^{\text{out}}_m \right)}_{\text{inference}},
\end{split}
\end{equation*}
where $n_e$ is the number of data events recorded, $d$ the total Lambda duration in gigabyte seconds, $n_w$ the DynamoDB write units consumed, and, for each model $m$ in the two model split of Section~\ref{sec:models}, $p^{\text{in}}_m$ and $p^{\text{out}}_m$ are the published per token prices and $T^{\text{in}}_m$, $T^{\text{out}}_m$ the token counts consumed. The EC2 host and the S3 storage for reports are fixed costs amortized across the day rather than charged per incident, so daily cost is
\[
c_{\text{day}} \;=\; \lambda\sigma\, c_{\text{inc}} \;+\; c_{\text{ec2}} \;+\; c_{\text{s3}} .
\]

The structure of these expressions carries the argument. In a deception anchored design $c_{\text{day}}$ is linear in $\lambda$, the decoy interaction rate, and independent of the total control plane event volume of the account. The infrastructure terms are charged only against events on the decoy, and the inference terms are charged only when a decoy session flushes. In an anomaly triggered design the invocation rate is instead set by the detector's alert rate, which grows with account activity and with the detector's false positive rate. Consider an account generating $10^6$ control plane events per day paired with a detector operating at a one in $10^4$ alert rate. That configuration produces on the order of $10^2$ agent invocations per day, and the per invocation cost is itself higher because an agent investigating an ambiguous anomaly must retrieve more context before it can reach a conclusion than an agent investigating a decoy touch, for which the anchor is unambiguous. The same account instrumented with a decoy produces invocations only when the decoy is touched, which for a resource no legitimate workflow references is a rate governed by adversary and scanner behavior rather than by the size of the environment. The arithmetic here is illustrative rather than measured, since we did not run an anomaly triggered comparison, and Section~\ref{sec:missing} lists that comparison among the ones this evaluation does not make.

This is the precise sense in which we claim deception changes the engineering economics of agentic investigation. It decouples the cost of investigation from the size of the environment being investigated, which is what makes an agent economically viable behind a decoy gate that would not be viable against the full baseline of legitimate activity.

% =====================================================================
\section{Discussion}
% =====================================================================
Three observations bear on whether the design generalizes beyond the validation environment.

The first concerns the role of deception as a selector rather than as a destination. In this framework the decoy does not exist in order to surface alerts to a SOC analyst; it exists in order to determine which sessions enter the agentic pipeline at all. The high signal to noise ratio of the decoy is what makes the downstream agent practical, and Section~\ref{sec:cost} gives that observation its quantitative form. We expect the same logic to apply to other deception primitives such as honey IAM roles, honey secrets in parameter stores, and honey functions, each of which would compose into the same downstream pipeline by exposing an additional MCP tool.

The second concerns dynamic prompt generation. We chose the two stage formulation because it binds reasoning constraints once and evidence per incident, but the deeper reason we believe it works is that it removes the temptation a static prompt creates to fill placeholder fields with plausible text. A template listing ten standard fields will be filled with ten standard looking values whether or not the evidence supports them, and the hallucination literature is clear that this is a primary failure mode for security applications \cite{huang2023halsurvey, alharthi2025llmforensics}. The dynamic prompt, by carrying only observed fields, removes the placeholder and therefore removes the failure mode. We hold this as a design argument supported by a suite in which no unsupported assertion appeared, not as an isolated experimental result, and we are explicit about the difference.

The third concerns the relationship between the agent's outputs and downstream automation. The report is human facing, but the structured representation the agent emits before rendering could drive automated remediation, including revoking the implicated access key, isolating the implicated identity, blocking the implicated source address, and rotating any secret that lived behind the decoy. We did not validate automated remediation, partly because the validation environment contains no production assets to remediate and partly because automation raises safety questions that warrant their own evaluation. The analysis in Section~\ref{sec:untrusted} bears directly on that future work, because an actuating pipeline driven by partly adversary authored telemetry raises the consequence of a successful injection from a misleading report to a misdirected remediation. Learned policies for constrained response under a zero trust assumption offer one direction \cite{alharthi2024zerotrust}.

% =====================================================================
\section{Limitations and Future Work}
\label{sec:limitations}
% =====================================================================
We separate what this evaluation does not establish from what the system does not yet do, because the two call for different remedies.

\subsection{Comparisons This Evaluation Does Not Make}
\label{sec:missing}
The evaluation reported here has no baseline and no ablation. It establishes that the system as built reconstructs nine of ten controlled scenarios and satisfies Invariant 1 across the suite, and it establishes nothing about which component is responsible. We name the specific comparisons required, in the order we consider them most informative.

A raw log baseline would pass the full CloudTrail record set for the trigger window to the report model under a fixed template prompt, with no session aggregation, no enrichment, and no dynamic assembly. This is the naive application of a language model to log analysis and it is the condition against which any claim of architectural contribution has to hold. A static template condition would apply session aggregation but replace the dynamic prompt with a fixed template carrying placeholder fields, which is the comparison that would isolate the prompt assembly discipline from everything upstream of it and which bears directly on the central claim of this paper. A non deception trigger condition would drive the same agent, tools, and prompt discipline from an anomaly based trigger over general account activity, which is the comparison that would test whether the decoy gate contributes anything beyond convenience and would supply the measured alert rate that Section~\ref{sec:cost} currently supplies by illustration. A deterministic renderer condition would render the evidence package through a template engine with no model in the path, bounding how much of the reconstruction quality is attributable to the model at all.

Beyond baselines, five factors warrant ablation: the decoy gate, the session aggregation, the dynamic prompt, the MCP enrichment layer, and cross session correlation once implemented. Each should be removed with the others held fixed over the full suite.

Evaluation scale and setting are limitations in their own right. Ten scenarios on one service, one provider, and one region, executed by the authors in a clean account, do not approximate a production environment. Real environments carry more identities, cross account access, automation traffic that reaches an unintended decoy by misconfiguration, and partial logging configurations that produce missing field cases the prototype has not encountered. Broader evaluation in noisier sandboxes and eventually in an operational pilot is the appropriate next validation step.

We did not release an implementation artifact with this work. Reproduction from the description in Sections~\ref{sec:formal} through~\ref{sec:impl} is possible but is not the same thing as reproduction from released code, and we regard this as a limitation rather than as a neutral choice.

\subsection{Cross Session Correlation}
\label{sec:futurecorr}
The delayed access failure of Section~\ref{sec:delayed} points at cross session correlation as the next architectural improvement, and we state the procedure here so that it is specified rather than gestured at. Algorithm~\ref{alg:xsession} is not implemented in the system evaluated in this paper.

After a session is flushed, the agent queries a correlation store for prior sessions within a correlation horizon $H \gg \tau$ whose pivots overlap on at least the identity or credential components, scores each candidate, and merges when the score exceeds a threshold $\theta$. The score combines four terms. Pivot overlap counts matching components of the pivot tuple, weighted so that a matching durable principal counts for more than a matching source address, since addresses are shared behind network address translation and principals are not. Temporal proximity decays exponentially in the inter session gap. Sequence consistency asks whether the concatenated event sequence is consistent with a single progression through the kill chain, so that an enumeration session followed by a retrieval session scores higher than two enumeration sessions. Resource continuity rewards overlap in the set of object keys touched, computed over hashed keys so that client supplied strings never enter the scoring logic in raw form.

\begin{algorithm}[t]
\caption{Cross session correlation (specified, not implemented)}
\label{alg:xsession}
\begin{algorithmic}[1]
\Require flushed session $S$; correlation store $\mathcal{S}$; horizon $H$; threshold $\theta$
\Ensure campaign $C$ containing $S$
\State $\mathcal{K} \gets \{S' \in \mathcal{S} : |t_{\min}(S) - t_{\max}(S')| \le H \}$
\State $\mathcal{K} \gets \{S' \in \mathcal{K} : \rho(S') = \rho(S) \lor \kappa(S') = \kappa(S)\}$
\State $C \gets \{S\}$
\ForAll{$S' \in \mathcal{K}$}
  \State $g \gets |t_{\min}(S) - t_{\max}(S')|$
  \State $s \gets w_1 \cdot \mathrm{ov}(S,S') + w_2 \cdot e^{-g/H}$
  \State \hspace{1.1em} $+\, w_3 \cdot \mathrm{seq}(S',S) + w_4 \cdot \mathrm{res}(S',S)$
  \If{$s \ge \theta$}
    \State $C \gets C \cup \mathrm{campaign}(S')$
  \EndIf
\EndFor
\State \Return $\mathrm{merge}(C)$ ordered by event time
\end{algorithmic}
\end{algorithm}

Indexing the correlation store by durable principal and by access key retrieves the candidate set in time logarithmic in the number of stored sessions, and the scoring loop is linear in that set. The candidate set is bounded by the number of decoy interactions within $H$, which is small by construction, so this is another place where the deception gate buys tractability that an unanchored pipeline would not have. Merging should be monotone and idempotent so that a late arriving session joins an existing campaign without rewriting it, and the merged campaign re rendered rather than appended to, which keeps the report a single coherent narrative rather than a stack of updates. Selecting $H$, $\theta$, and the four weights, and measuring the resulting behavior on delayed and concurrent actor scenarios, is the immediate next piece of work.

\subsection{Injection Resistance}
\label{sec:futureinjection}
Section~\ref{sec:untrusted} identifies cloud telemetry as a partially adversary authored channel and Section~\ref{sec:formal} carries the partition into the pivot definition, which is the extent of the mitigation the current prototype implements. Closing the remainder requires four changes. Client supplied fields should be length truncated and escaped at normalization, with control characters and prompt delimiter sequences neutralized. Correlation and scoring logic should read those fields only through a salted hash, never in raw form. The user prompt should carry them inside an explicitly delimited region that the system prompt identifies as data rather than direction. And the report renderer should escape them on output so that a crafted object key cannot alter the structure of the rendered document.

Evaluating these requires a probe suite in which instruction text is placed in object keys and user agent strings, measuring three things separately: whether the crafted string appears verbatim in the report, which it should, since suppressing evidence is its own failure; whether the model acts on it; and whether any unintended tool call fires. We regard this as the highest priority extension, because a deception anchored pipeline is the configuration in which an adversary has the most direct write access to the analysis input.

\subsection{Decoy Fingerprinting and Evasion}
The most consequential limitation is the one the design cannot argue away. An adversary who identifies the decoy and avoids it produces no signal, and the entire pipeline is silent. We did not evaluate detectability, and we note that an authenticated adversary has access to signals that would plausibly separate a decoy from production resources in the same account, including creation timestamp clustering, object count and size distributions, access frequency by legitimate principals, naming entropy relative to sibling resources, and configuration divergence in versioning, lifecycle, and tagging. Measuring which of these separate a decoy in practice, and which admit mitigation, is necessary work that this paper does not do.

The deception literature has explored detection and evasion at length without settling the question for cloud native substrates \cite{javadpour2024deception}. Our operative guidance is the combination of multiple decoy surfaces so that avoidance must be complete rather than partial, naming and configuration drawn from the tenant's own conventions rather than from a template, and rotation of decoy contents on a schedule that outpaces reconnaissance. We do not claim this makes a decoy undetectable. We claim it raises the cost of complete avoidance above the cost of partial avoidance, which is the achievable goal.

\subsection{Surface and Provider Generalization}
The validation focused on Amazon S3 as the decoy surface. S3 was a natural starting point because its data event coverage is rich, its security context is well understood, and prior measurement work established that scanners reach S3 buckets in the wild \cite{izhikevich2024honeybuckets}. Extending to Secrets Manager entries, Lambda functions, RDS instances, and IAM roles is the clear next step, and each requires its own context tool and its own entries in the timeline classifier. Multi cloud generalization is a related axis. We expect the agent layer to port because MCP tools are vendor agnostic by construction, though the pivot tuple would need redefinition for a provider whose identity model differs from the IAM and STS arrangement, and that redefinition is where the difficulty would concentrate.

Within the agent, three enhancements would strengthen trust. Per claim evidence citations would make Invariant 1 auditable by the reader rather than only by the authors, and recent work on provenance native answer traces provides both the measurement framework and the leave one resource out intervention needed to verify that a cited source actually shaped the claim \cite{faizan2026provenai}. Confidence scoring derived from the consistency of overlapping tool outputs would let the report distinguish well supported claims from thinly supported ones. And a stricter schema contract between the timeline stage and the report stage would move part of Invariant 1 from review into type checking.

% =====================================================================
\section{Conclusion}
% =====================================================================
Cloud incident understanding is bottlenecked less by the absence of evidence than by the cost of reconstructing it under time pressure. Cloud Decoy AI Agent attacks that bottleneck from two sides. On the upstream side it uses a cloud native decoy as the entry point so that every session entering the pipeline is by construction worth investigating, and Section~\ref{sec:cost} shows that this decouples investigation cost from environment size. On the downstream side it uses an autonomous tool using agent under a stated grounding invariant to compress the path from a raw session to an analyst ready report. Across ten controlled scenarios the system reconstructed nine completely, produced no assertion the authors could not trace to an observed artifact, and delivered reports in four to five minutes.

Three elements are, in our reading, the ones most likely to generalize beyond this setting. The first is the treatment of the deception signal as a cost and correctness bound on an agent's evidence horizon rather than as an alert to be queued. The second is the two stage prompt construction that enforces evidence boundedness structurally rather than by instruction. The third is the recognition that cloud telemetry is partly adversary authored, which makes any log to prompt pipeline an injection surface and which the deception setting widens rather than narrows, and which we identify here without yet mitigating.

We are equally clear about what this work does not settle. The evaluation has no baseline, no ablation, and no independent scoring, and it covers one service on one provider. Section~\ref{sec:missing} names the specific comparisons that would resolve each of these, and we regard them as the necessary next step rather than as optional refinement. We expect deception driven agentic investigation to become a common pattern in cloud incident response, and we offer this work as an early and openly bounded instance of the pattern.

%\section*{Acknowledgment}
%The authors thank the University of Arizona College of Information Science for supporting this work, and the anonymous reviewers of an earlier version of this manuscript, whose criticism shaped the threat model, the cost analysis, and the account of this evaluation's limits presented here.

\bibliographystyle{IEEEtran}
\bibliography{references-46}

\appendices

% =====================================================================
\section{Scoring Rubric}
\label{app:rubric}
% =====================================================================
The criteria below were applied to each generated report against the driver script ground truth. We reproduce them so that the judgments reported in Section~\ref{sec:results} are auditable in form, while noting again that the review was performed by the authors and that an independent application of the same rubric would carry materially more weight.

For event coverage, each ground truth API call is marked present or absent in the reconstructed timeline, and the scenario passes only when every call is present. For ordering, each consecutive pair of ground truth events is marked correctly or incorrectly ordered, and the scenario passes only when every pair is correct. For attribution, the acting credential, the durable principal, and the role chain where one exists are recorded separately, and the scenario passes only when all applicable components are correct. For object naming, each accessed object key is marked named or not named, and the scenario passes only when every key is named; scenarios without retrieval are excluded from this count. For unsupported assertions, every factual assertion in the report is extracted, where a factual assertion is any statement naming an identifier, a timestamp, an object, an event, a count, or a byte quantity, and each is marked supported or unsupported against the evidence package; the scenario passes only when no unsupported assertion appears.

A scenario receives a score of $1$ when it passes every criterion, and $0.5$ when it passes most but fails one non trivial criterion. No score below $0.5$ was awarded, and none was warranted in this suite.

\balance

\end{document}